\documentstyle[12pt]{article}
\newcommand{\lsim}   {\mathrel{\mathop{\kern 0pt \rlap
  {\raise.2ex\hbox{$<$}}}
  \lower.9ex\hbox{\kern-.190em $\sim$}}}
\newcommand{\gsim}   {\mathrel{\mathop{\kern 0pt \rlap
  {\raise.2ex\hbox{$>$}}}
  \lower.9ex\hbox{\kern-.190em $\sim$}}}
\newcommand{\K}{\textrm{K}}

\newcommand{\Od}{{\cal O}}

\newcommand{\mapright}[1]{\smash{\mathop{\hbox to 1cm{\rightarrowfill}}\limits_{#1}}}
\def\gappeq{\mathrel{\rlap {\raise.5ex\hbox{$>$}}
{\lower.5ex\hbox{$\sim$}}}}
\def\lappeq{\mathrel{\rlap{\raise.5ex\hbox{$<$}}
{\lower.5ex\hbox{$\sim$}}}}

\begin{document}
\input epsf
\pagestyle{empty}
\begin{flushright}
\end{flushright}
\vspace*{5mm}
\begin{center}
\Large{\bf Cosmology with moving dark energy and the
CMB quadrupole} \\
\vspace*{1cm}
\large{\bf Jos\'e Beltr\'an Jim\'enez$^1$}  and
\large{\bf Antonio L. Maroto$^2$} \\
\vspace{0.3cm}
\normalsize
Departamento de F\'{\i}sica Te\'orica\\
Universidad Complutense de Madrid\\
28040 Madrid, Spain

\vspace*{1cm}
{\bf ABSTRACT} \\ \end{center}
We study the consequences of a homogeneous dark energy fluid
having a non-vanishing velocity with respect
to the matter and radiation large-scale rest frames. We consider
 homogeneous  anisotropic cosmological models with four
fluids (baryons, radiation,
dark matter and dark energy) whose velocities can differ
from each other. Performing a perturbative calculation up to second
order in the velocities, we obtain the contribution of the anisotropies
generated by the fluids motion to the CMB quadrupole and compare
with observations. We also consider
the exact problem for arbitrary velocities and solve the corresponding
equations numerically
for different dark energy models.
We find that models whose equation of state is initially stiffer than
radiation,
as for instance some tracking models, are unstable
against velocity perturbations, thus spoiling the late-time predictions
for the energy densities. In the case of scaling models, the contributions
to the quadrupole can be non-negligible for a wide range of initial
conditions. We also consider fluids
moving at the speed of light (null fluids) with positive energy
and show that, without
assuming any particular equation of state,
they generically act as a cosmological constant at late times.
 We find the parameter region for which the models considered could
be compatible with the measured (low) quadrupole.

\vspace*{5mm}
\noindent

\vspace*{0.5cm}


\vspace{0.5cm}

\noindent
\rule[.1in]{8cm}{.002in}

\noindent $^1$E-mail: jobeltra@fis.ucm.es\\
\noindent $^2$E-mail: maroto@fis.ucm.es
\vfill\eject

\setcounter{page}{1}
\pagestyle{plain}

\newpage
\section{Introduction}

The present observational evidence from supernovae type Ia
\cite{Gold,SNLS}, CMB anisotropies \cite{WMAP3}
and large scale structure, mainly through the baryon acoustic
oscillations, suggests that today the universe could be
dominated by a negative pressure fluid \cite{review,review2}. Although these data are
compatible with the presence of a cosmological constant, the fact that
such observations only explore relatively recent epochs
implies that other possible models in which the equation
of state of dark energy could have changed in time  cannot be
excluded a priori. Thus, if dark energy can be parametrized as
a perfect fluid with equation of state $p_{DE}=w_{DE}(z)\rho_{DE}$,
the above mentioned observations only constrain the present value as
$w_{DE}^0=-0.97^{+0.07}_{-0.09}$ \cite{WMAP3}. The redshift dependence
of the equation of state
can be parametrized in different ways
and still a wide range of variability is compatible with observations
\cite{Periv}. Moreover, apart from including new components in the energy-momentum tensor
within Einstein gravity, or new scalar fields as in the quintessence models
\cite{quintessence},
it has been also suggested that the
accelerated expansion
of the universe could be due to modifications of the gravitational action at
large distances \cite{infrared}.

The possibility of finding  observational signals of
dark energy which
could discriminate between the various models is thus
becoming of crucial importance. However, up to date, only a
few proposals have been
considered in the literature, which can be broadly classified in two
classes \cite{Trotta}: on one hand probes of the redshift-distance
relation, as for instance the already mentioned high-redshift supernovae
Ia or the use of baryon acoustic oscillations as standard rulers.
On the other hand we have the probes of the growth of structure in the
universe, such as the weak gravitational lensing or
the integrated Sachs-Wolfe effect.
The goals of the future observational surveys will be the
determination of the equation of state of dark energy with a
few percent accuracy and the possibility of discriminating from
a pure cosmological constant \cite{task}.

In this paper we consider a  different aspect
of dark energy with potential observational
consequences which is related to the possible motion
of dark energy with respect to matter and radiation.
If dark energy can be described  as a homogeneous perfect
fluid then, apart from the density parameter $\Omega_{DE}$ and
equation of state $w_{DE}$, a complete knowledge of its energy-momentum
tensor requires the determination of its relative velocity with
respect to the rest of  components of the universe.
Indeed, dark energy is usually considered as a highly
homogeneous fluid in a similar way to radiation. This is
due to the fact that in most models, its sound speed is
close to the speed of light and this prevents the growth
of dark energy perturbations below the Hubble scale.
Moreover, dark energy
is also required to be extremely weakly interacting with baryons and
radiation, in order to avoid conflicts with the predictions of
Standard Cosmology. Indeed, in most of the models, its effects are
purely gravitational and dark energy is considered as a totally decoupled
fluid. In such a case it makes sense to ask whether the dark energy
rest frame converges towards the radiation or matter large-scale rest frames.
Since there is no a priori reason to expect that dark energy were necessarily
coupled to radiation in the very early universe,
its initial velocity  with respect to radiation should
be considered as a free cosmological parameter on equal footing to
$\Omega_{DE}$ or $w_{DE}$, and very much in the
same way as those  parameters, the relative velocity of
dark energy  should be determined by observations. In addition,
the fact that a pure cosmological constant is invariant under
change of frame implies that the  potential effects associated
to a non-vanishing relative velocity will be exclusively present
in models with varying equation of state.

The metric anisotropies generated by the fluids motion can affect both
the temperature and polarization of the CMB.
In a previous work \cite{maroto},  we have started the study 
of such effects on the
CMB dipole.
We have shown that the motion of dark energy is not
incompatible with the current measurements of the dipole, but instead, it
 modifies  its usual interpretation. Thus, 
 when dark energy is moving, the dipole is generated by the motion
of emitter and observer  with
respect to the cosmic center of mass and not with respect 
to the background radiation. This fact can also have important consequences
for the generation of matter bulk flows on very large scales
\cite{rest}.
In
this work we consider the effects of motion on the CMB quadrupole.
 In the last years, the interest in anisotropic models 
(see \cite{Barrow,inflationfluctuation} and references therein) 
has grown mainly motivated by some unexpected features in the low
multipoles of the CMB temperature anisotropies, in particular
the low value of the quadrupole and the quadrupole-octupole alignment 
which could
suggest the existence of a preferred direction in the universe
\cite{Schwarz,Magueijo}. In this work we explore this possibility by considering
cosmologies with several moving fluids (see
\cite{2fluids} for previous works).

The paper is organized as follows: in Section 2 we obtain the metric
solution for a model with moving fluids up to second order in the velocities.
In Section 3, we obtain the corresponding CMB temperature anisotropies
and the contribution to the quadrupole. In  Section 4 we consider the
exact problem for large velocities and obtain the general expression for
the quadrupole. Section 5 is devoted to the application of the
previous results to different dark energy models. Finally Section 6
contains the main conclusions of the paper.

\section{Slow-moving fluids: second order equations}
Let us consider a universe filled with four homogeneous perfect
fluids: baryons, radiation, dark matter and dark energy. For
matter and radiation we shall consider their usual equations of
state, i.e., $p_R=\frac{1}{3}\rho_R$ and $p_B=p_{DM}=0$, whereas
for dark energy we shall use: $p_{DE}=w_{DE}(z)\rho_{DE}$, which in
general depends on redshift. In an arbitrary frame, the
energy-momentum tensor for each component takes the  perfect
fluid form:
\begin{eqnarray}
T^{\mu\nu}_\alpha=(\rho_\alpha+p_\alpha)u_\alpha^\mu
u_\alpha^\nu-p_\alpha g^{\mu\nu}, \label{emtensor}
\end{eqnarray}
where $\alpha=B,R, DM, DE$. We shall consider homogeneity in the
fluids so that all the quantities appearing in (\ref{emtensor}) will
just depend on conformal time:
\begin{eqnarray}
\rho_\alpha&=&\rho_\alpha(\eta),\nonumber\\
u^\mu_\alpha&=&\gamma_\alpha(\eta)\left(1,\vec
v_\alpha(\eta)\right)
\end{eqnarray}
with
\begin{eqnarray}
\gamma_\alpha=\frac{1}{\sqrt{g_{00}+g_{ij}v_\alpha^iv_\alpha^j}}.
\end{eqnarray}
In order to simplify the problem and to obtain some analytical
solutions we shall use perturbation theory, assuming
that the fluids velocities are
small, i.e.  $\vec v_\alpha^2\ll 1$. To that end, we
expand the different quantities of the four fluids up to second order as
follows:
\begin{eqnarray}
\rho_\alpha&=&\rho^{(0)}_\alpha+\rho^{(1)}_\alpha+\rho^{(2)}_\alpha+\cdot
\cdot \cdot\nonumber\\\vec{v}_\alpha
&=&\vec{v}^{(1)}_\alpha+\vec{v}^{(2)}_\alpha+\cdot \cdot \cdot
\end{eqnarray}
where we have imposed  the fluids to be at rest to zeroth order,
i.e., $\vec{v}^{(0)}_\alpha=0$. That way, the most general form for
the metric is given by the perturbed Friedmann-Robertson-Walker
metric:
\begin{eqnarray}
ds^2&=&a^2\left[\left[1+2\left(\phi^{(1)}+\phi^{(2)}\right)\right]d\eta^2+2\left[S^{(1)}_i+S^{(2)}_i\right]dx^id\eta-\right.\nonumber\\
&&\left.\left[\left(1-2\left(\psi^{(1)}+\psi^{(2)}\right)\right)\delta_{ij}+h_{ij}\right]dx^idx^j\right]
\end{eqnarray}
where we will follow the notation in \cite{Mukhanov}.
In this expression, $\phi^{(1)}$ and $\psi^{(1)}$ are scalar
perturbations of first order and will be determined from
$\rho^{(1)}_\alpha$ in the first order equations of motion.
However, the second order scalar perturbations $\phi^{(2)}$ and
$\psi^{(2)}$ can depend, not only on $\rho^{(2)}_\alpha$ and
$(\rho^{(1)}_\alpha)^2$ terms, but also on
$(\vec{v}^{\;(1)}_\alpha)^2$ which are also scalars. Analogously,
the first order vector perturbations $\vec{S}^{(1)}$ can only be
related to the first order velocities $\vec{v}^{(1)}_\alpha$ in
the equations of motion, whereas to second order, $\vec{S}^{(2)}$
will be determined by combinations of $\vec{v}^{(2)}_\alpha$ and
$F^{(1)}\vec{v}^{(1)}_\alpha$ where $F^{(1)}$ is a scalar function
of the first order scalar perturbations. Finally, $h_{ij}$ is a
traceless tensor perturbation which should be of second order and
depend on the combinations $v^{(1)}_{\alpha i}v^{(1)}_{\alpha
j}-\frac{1}{3}(v^{(1)}_\alpha)^2\delta_{ij}$. Note that, since we
are considering only time-dependent perturbations on the
energy-momentum tensor, all the perturbations on the metric will be
functions only of time and, therefore, the perturbed metric does
not contain any terms involving spatial derivatives.

So far, we have not done any specific gauge choice, so we still have four
gauge degrees of freedom. Hence, we can simplify the problem by chosing
our coordinates appropriately. In particular, we can
fix the spatial coordinates so that the vector part of the metric
vanishes $\vec S=0$. The physical interpretation of this
condition is apparent when solving the  $(^0_{\;\;\;i})$ Einstein equation
of the exact problem. Thus, we obtain
the condition:
\begin{eqnarray}
S_i=\frac{\sum_\alpha\gamma_\alpha^2(\rho_\alpha+p_\alpha)g_{ij}v^j_\alpha}
{\sum_\alpha\gamma_\alpha^2(\rho_\alpha+p_\alpha)},\label{ccmc}
\end{eqnarray}
where the metric is $ds^2=g_{\mu\nu}dx^\mu dx^\nu$. Hence,
$\vec{S}$ can be interpreted as the relativistic cosmic center of
mass velocity (see \cite{maroto}). Notice
that in general, an observer at rest with respect to cosmic
center of mass could be moving with respect to radiation or
matter.
On the other
hand, the temporal coordinate can be chosen in such a way that
$\sum_\alpha(\rho_\alpha-\rho^{(0)}_\alpha)=0$, which means that the total
density perturbations are identically zero. With this gauge choice,
the $(^0_{\;\;\;0})$ and $(^i_{\;\;\;j})$ components of
Einstein equations $G^\mu_{\;\;\nu}=8\pi GT^\mu_{\;\;\nu}$ up
to second order adopt the form:

{\bf Zeroth order}
\begin{eqnarray}
{\cal H}^2&=&\frac{8\pi G}{3}\,a^2\sum_\alpha \rho^{(0)}_\alpha \\
2{\cal H}'+{\cal H}^2&=&-8\pi G\,a^2\sum_\alpha p^{(0)}_\alpha
\end{eqnarray}

{\bf First order}
\begin{eqnarray}
-\frac{6}{a^2}{\cal H}\left(\psi'^{(1)}+{\cal H}\phi^{(1)}\right)=0
\end{eqnarray}
\begin{eqnarray}
\psi''^{(1)}+2{\cal H}\psi'^{(1)}+{\cal H}\phi'^{(1)}+({\cal
H}^2+2{\cal H}')\phi^{(1)}=0
\end{eqnarray}

{\bf Second order}
\begin{eqnarray}
&-&2{\cal{H}}\left(\psi^{(2)}+\left(\psi^{(1)}\right)^2\right)'-
2{\cal{H}}^2\left(\phi^{(2)}-2\left(\phi^{(1)}\right)^2\right)+
\psi'^{(1)}\left(\psi'^{(1)}+4{\cal{H}}\phi^{(1)}\right)
\nonumber\\
&=&\frac{8\pi
G}{3}\,a^2\sum_\alpha\left(\rho^{(0)}_\alpha+p^{(0)}_\alpha\right)
\left(v^{(1)}_\alpha\right)^2\\\nonumber\\
&\left.\right.&\frac{2}{a^2}\left[\left(2{\cal{H}}'+{\cal{H}}^2\right)\phi
^{(2)}+{\cal{H}}\phi'^{(2)}+\psi''^{(2)}+2{\cal{H}}\psi'^{(2)}
-2\left({\cal{H}}^2+2{\cal{H}}'\right)\left(\phi
^{(1)}\right)^2\right.\nonumber\\
&+&\left.\frac{1}{2}\psi'^{(1)}\left(\psi^{(1)}
-2\phi^{(1)}\right)'+
2\psi''^{(1)}\left(\psi^{(1)}-\phi^{(1)}\right)+2{\cal{H}}
\left[\left((\psi^{(1)})^2+(\phi^{(1)})^2\right)'+2\phi^{(1)}
\psi'^{(1)}\right]\right]\delta^i_j
\nonumber\\
&+&\frac{1}{2a^4}
\left(a^2h'^{\,i}_{\;\;j}\right)'
=8\pi
G\sum_\alpha\left(\rho^{(0)}_\alpha+p^{(0)}_\alpha\right)v^{(1)i}_\alpha
v^{(1)}_{\alpha j} \label{Eeh}
\end{eqnarray}
with $'\equiv\frac{d}{d\eta}$ and ${\cal H}=a'/a$ is the Hubble
parameter.

Nevertheless, the system is incomplete because there are more
unknown variables than equations. In general, the problem with $n$
fluids has ten independent Einstein equations, but the unknown
quantities are the densities (assuming a given equation of state)
and the three independent components of the four-velocity of each
fluid (because of the constraint $u^2=1$). Therefore, there are
$10+4n$ unknown functions,
although, since there are four gauge degrees of freedom, we can
fix four quantities and reduce the number of undetermined
functions to $6+4n$. With this count, one  needs $4(n-1)$
additional equations to complete the system. The simplest way to
close the problem is by requiring the conservation of each
energy-momentum tensor, assuming they are decoupled from each other:
$T^{\mu\nu}_{\alpha\;\; ;\nu}=0$. Obviously,
one can modify these relations by changing the right hand side in
order to consider interactions between the fluids.
This guarantees the completeness of the system since it provides
the $4(n-1)$ required equations (there are $4n$ extra equations,
but the conservation of the total energy-momentum tensor makes one
of those equations superfluous). For our case, these additional
equations read for the energy and momentum conservation (notice
that momentum conservation is trivial at zeroth order):

{\bf Zeroth order}
\begin{eqnarray}
\rho_\alpha'^{(0)}+3{\cal
H}\left(\rho^{(0)}_\alpha+p^{(0)}_\alpha\right)=0\label{ece0}
\end{eqnarray}

{\bf First order}
\begin{eqnarray}
\rho_\alpha'^{(1)}+3{\cal H}\left(\rho^{(1)}_\alpha+p^{(1)}_\alpha\right)&=&3\left(\rho^{(0)}_\alpha+p^{(0)}_\alpha\right)\psi'^{(1)}\\
\left[a^4\left(\rho^{(0)}_\alpha+p^{(0)}_\alpha\right)
\vec{v}^{\;(1)}_\alpha\right]'&=&0\label{ecv1}
\end{eqnarray}

{\bf Second order}

\begin{eqnarray}
\rho_\alpha'^{(2)}+3{\cal H}(\rho^{(2)}_\alpha+p^{(2)}_\alpha)&=&
-\frac{1}{a^4}\left[a^4\left(\rho^{(0)}_\alpha+p^{(0)}_\alpha\right)\right]'
\left(\vec{v}^{\,(1)}_\alpha\right)^2-(\rho^{(0)}_\alpha+p^{(0)}_\alpha)
\left((\vec{v}^{\;(1)}_\alpha)^2\right)'\nonumber\\
&+&3\left[\left(\rho^{(0)}_\alpha+p^{(0)}_\alpha\right)
\left((\psi^{(1)})^2+\psi^{(2)}\right)'
+(\rho^{(1)}_\alpha+p^{(1)}_\alpha)\psi'^{(1)}\right]
\end{eqnarray}
\begin{eqnarray}
\left[a^4\left((\rho^{(0)}_\alpha+p^{(0)}_\alpha)
(\vec{v}^{(2)}_\alpha-2\phi^{(1)}\vec{v}^{(1)}_\alpha)+(\rho^{(1)}_\alpha
+p^{(1)}_\alpha)
\vec{v}^{(1)}_\alpha\right)\right]'=a^4(\rho^{(0)}_\alpha+p^{(0)}_\alpha)
(5\psi^{(1)}-\phi^{(1)})'\vec{v}^{(1)}_\alpha
\end{eqnarray}
where we have used the previous orders equations at each order.

\section{Contributions to the CMB quadrupole}

The relevant part of the metric for the quadrupole is
given in our case just by $h_{ij}$, since it is the only part
contributing to  the anisotropy.   The homogeneous
scalar perturbations only affect the value of the monopole
in an negligible way. Concerning  vector perturbations, we are working
in  the cosmic center of mass frame and accordingly those contributions
vanish in our calculations. As commented in the introduction,
the $\vec S$ contributions  have been
studied in \cite{maroto} and modify
the usual dipole contribution. Therefore, in order to calculate
the quadrupole produced
by this metric, we can consider just the tensor perturbation.
Then, from now on, we shall use the metric:
\begin{eqnarray}
ds^2=a^2(d\eta^2-(\delta_{ij}+h_{ij})dx^idx^j).\label{mt}
\end{eqnarray}
In order to calculate all the contributions to the temperature anisotropies
generated by the metric perturbations, 
we should solve the corresponding radiative transfer equations (see
\cite{Dodelson,Gio}). This is the system of Einstein-Boltzmann equations
for the set of fluids. However, since we are only interested in the 
quadrupole anisotropy (which is not affected by microphysics at the time
of recombination), the only relevant contribution for such a large 
angle contribution  would be given by the Sachs-Wolfe effect 
which takes into account the variation in the energy of photons propagating
from the last scattering surface \cite{Gio}:
\begin{eqnarray}
\frac{\delta T}{T}=\frac{a_0{\cal E}_0-a_{dec}{\cal
E}_{dec}}{a_{dec}{\cal E}_{dec}}. \label{SW}
\end{eqnarray}
Here, the indices $0$ and $dec$ denote the present and decoupling
times respectively and ${\cal E}$ is the energy of the photon. For
an observer with velocity $u^\mu=\gamma(1,\vec{v})$ this energy is
given by:
\begin{eqnarray}
{\cal E}=g_{\mu\nu}u^\mu P^\nu, \label{energy}
\end{eqnarray}
with
\begin{eqnarray}
P^\nu=E\frac{dx^\nu}{d\lambda},
\end{eqnarray}
where $E$ parametrizes the photon energy and $\lambda$ is an
affine parameter. By the invariance of the action of the geodesics
of a massless particle under conformal transformations of the
affine parameter, the geodesics of the metric $g_{\mu\nu}$ given
by (\ref{mt}) with affine parameter $\lambda$ are the same as
those of the metric $\hat{g}_{\mu\nu}=a^{-2}g_{\mu\nu}$ with
affine parameter $\eta$ such that $d\lambda=a^2d\eta$. The
trajectory of the photon coming from the direction given by the
Minkowski-null vector $n^\mu=(1,\vec{n})$ with $\vec{n}^2=1$
will be perturbed  in such a way that we can write
$x^\mu(\eta)=n^\mu\eta+\delta x^\mu$, where the second term
corresponds to the contribution from $h_{ij}$ which is of second
order. Then,
assuming that the observer velocity is of first order, the
momentum of the photon to second order is:
\begin{eqnarray}
P^\nu=\frac{E}{a^2}\left(n^\nu+\frac{d\delta
x^\nu}{d\eta}\right).
\end{eqnarray}
Inserting this expression in (\ref{energy}) we obtain:
\begin{eqnarray}
{\cal E}=\frac{E}{a}\left(1+\frac{1}{2}\vec{v}^{\,
2}-\vec{v}\cdot\vec{n}+\frac{d\delta x^0}{d\eta}\right).
\end{eqnarray}
For the $\hat{g}_{\mu\nu}$ metric, the second order of the zero
component of the geodesic equations in terms of the metric
perturbation reduces to:
\begin{eqnarray}
\frac{d^2\delta
x^0}{d\eta^2}+\frac{1}{2}\frac{dh_{ij}}{d\eta}n^in^j=0
\end{eqnarray}
which can be easily integrated to get:
\begin{eqnarray}
\frac{d\delta x^0}{d\eta}=-\frac{1}{2}h_{ij}n^in^j.
\end{eqnarray}
Then, the energy of the photon results finally:
\begin{eqnarray}
{\cal E}=\frac{E}{a}\left(1+\frac{1}{2}\vec{v}^{\,
2}-\vec{v}\cdot\vec{n}-\frac{1}{2}h_{ij}n^in^j\right).
\end{eqnarray}
Thus, by using this formula for the energy of the photon in
equation (\ref{SW}) and expanding up to second order we obtain the
following expression for the temperature fluctuations:
\begin{eqnarray}
\frac{\delta
T}{T}\simeq\frac{1}{2}\vec{v}^{\;2}|^0_{dec}-\vec{v}\cdot\vec{n}|^0_{dec}-(\vec{v}_{dec}\cdot\vec{n})(\vec{v}\cdot\vec{n})|^0_{dec}-\frac{1}{2}h_{ij}
n^in^j |_{dec}^0. \label{Tf}
\end{eqnarray}
The first term in (\ref{Tf}) only contributes to the monopole and
the second term is a Doppler effect, although notice that since the
velocities appearing in $\vec{v}\cdot\vec{n}|^0_{dec}$ are
referred to the $\vec S=0$ frame, in the case of moving fluids,
 the dipole is
due to the motion of emitter and observer with respect to the
cosmic center of mass, as commented above.
Finally, the last two terms do contribute to the
quadrupole. The first of them does not depend
on the metric perturbation since it comes from the second order
expansion of the denominator in (\ref{SW}). As we will show below,
this term is expected to be smaller
than the last one. Therefore, the dominant
contribution for the quadrupole is given by the following expression:
\begin{eqnarray}
\frac{\delta
T_Q}{T}=-\frac{1}{2}(h_{ij}(a_0)-h_{ij}(a_{dec}))n^in^j.\label{quadrupole}
\end{eqnarray}
This  formula shows that we only need to know
$h_{ij}$ in order to calculate the quadrupole and, besides, this term 
does not depend on the observer velocity.
Moreover, it is easy to see from (\ref{Eeh}) that the
solution for $h_{ij}$ is the following:
\begin{eqnarray}
h_{ij}=6\int_{a_*}^{a}\frac{1}{\tilde a^4}\left[\int_{a_*}^{\tilde a}
\hat a^2\sum_\alpha\left(\rho^{(0)}_\alpha+p^{(0)}_\alpha\right)
\left(v_{\alpha
i}^{(1)}v_{\alpha
j}^{(1)}-\frac{1}{3}\vec{v}_\alpha^{(1)\,2}\delta_{ij}\right)\frac{d\hat a}
{\sqrt{\sum_\alpha\rho^{(0)}_\alpha}}\right]\frac{d\tilde a}
{\sqrt{\sum_\alpha\rho^{(0)}_\alpha}}\label{hij}
\end{eqnarray}
where $a_*$ is the  value of the scale factor at the time at
which we specify the initial conditions for
$h_{ij}$. Notice that the quadrupole does not depend on $h_{ij}(a_*)$,
but only on the initial value of the derivatives. For simplicity we
will assume that the metric anisotropies are generated by the fluids motion
 and therefore we consider a purely isotropic
universe for $a<a_*$, i.e., we take $h_{ij}(a_*)=h_{ij}'(a_*)=0$.

As we can see in the last expression, the quadrupole depends on
both the zeroth order densities of the fluids and the first order of
the velocities with respect to the center of mass. These
quantities can be obtained from the conservation equations
(\ref{ece0}) and (\ref{ecv1}). The zeroth order equations have the
well-known solutions for the densities $\rho_\alpha
^{(0)}=\rho_{0\alpha}a^{-3(w_\alpha+1)}$, $\rho_{0\alpha}$ being 
the densities today. On the other hand, this expression for the
zeroth order densities allows us to obtain the solutions for the
velocities:
\begin{equation}
\vec{v}_\alpha^{\;(1)}=\vec{v}_{0\alpha}a^{3w_\alpha-1}\label{vevol}
\end{equation}
where $\vec{v}_{0\alpha}$ are the velocities of the fluids today.
We see that, to first order, each fluid will move
along a fixed direction. Well inside the radiation dominated era,
baryons and dark matter particles were coupled
to radiation which, being  the dominant component, will drag
 matter in such a
way that the three fluids velocities are the same. From decoupling on,
each fluid velocity will start evolving according to (\ref{vevol}), i.e.
matter will start reducing its velocity with respect to the
cosmic center of mass frame, whereas radiation keeps moving at a
constant velocity. Since dark matter is expected to decouple before
baryons do, and both velocities scale in the same way, the two matter fluids
are expected to be moving with constant relative velocity after
recombination. Finally since the initial velocity of the three fluids
(radiation, baryons and dark matter) were the same, the direction of
their velocities will also agree after recombination.  On the
other hand, the gauge condition $\vec{S}=0$ in (\ref{ccmc})
yields the constraint:
\begin{equation}
\sum_\alpha\left(\rho_\alpha^{(0)}+p_\alpha^{(0)}\right)\vec{v}^{\,(1)}_\alpha=0\label{gaugecondition}
\end{equation}
so we can conclude that dark energy should also move in the same
direction as the rest of fluids in this frame. That way, each
energy-momentum tensor (and therefore the total one) will have
axisymmetry so the metric will also be axisymmetric. This means
that the tensor perturbation $h_{ij}$ is diagonal. In fact, if we
choose the velocities along the $z$-axis, the tensor perturbation
given by (\ref{hij}) will be proportional to $diag(-1,-1,2)$.
Taking into account the previous discussion, the final expression
for the quadrupole results:
\begin{eqnarray}
\frac{\delta
T_Q}{T}=-\frac{1}{2}\left(h(a_0)-h(a_{dec})\right)
\left(\cos^2\theta-\frac{1}{3}\right)\label{qexp}
\end{eqnarray}
where $\theta$ is the angle formed by the observation
direction and the velocities of the fluids, and
$h(a)=\sum_\alpha h_\alpha(a)$ with
\begin{eqnarray}
h_\alpha(a)=6\int_{a_*}^a\frac{1}{\tilde a^4}\left[\int_{a_*}^{\tilde a}
\hat a^2(\rho^{(0)}_\alpha+p^{(0)}_\alpha)v_{\alpha
}^{(1)\,2}\frac{d\hat a}{\sqrt{\sum_\alpha\rho^{(0)}_\alpha}}\right]\frac{d\tilde a}
{\sqrt{\sum_\alpha\rho^{(0)}_\alpha}}.\label{halpha}
\end{eqnarray}
The function of $\theta$ appearing in (\ref{qexp}) is proportional
to the spherical harmonic $Y_{20}$ so we can express the
quadrupole fluctuation as:
\begin{equation}
\frac{\delta
T_Q}{T}=\frac{2}{3}\sqrt{\frac{\pi}{5}}(h_0-h_{dec})Y_{20}.
\end{equation}
It is usual to introduce the power spectrum of the temperature
fluctuations of CMB as:
\begin{equation}
\frac{\delta
T_\ell}{T}=\sqrt{\frac{1}{2\pi}\frac{\ell(\ell+1)}{2\ell+1}\sum_m|a_{\ell
m}|^2}
\end{equation}
where $a_{\ell m}$ are the coefficients of the expansion in
spherical harmonics. Moreover, the quadrupole is usually defined
as:
\begin{equation}
Q\equiv\frac{\delta T_2}{T}=\sqrt{\frac{3}{5\pi}\sum_{m=-2}^2|a_{2
m}|^2}\label{Qdefinition}
\end{equation}
which, in our case, reduces to:
\begin{equation}
Q_A=\frac{2}{5\sqrt{3}}\left|h_0-h_{dec}\right|.\label{qfe}
\end{equation}

The
quadrupole given by (\ref{qfe}) is due to the anisotropy of the
space-time background (that is why the index $A$ is introduced),
but we have to
add the standard isotropic fluctuation produced during inflation.
Then, if we assume that the anisotropies are small, the total
effect will be the linear superposition of both contributions (see
\cite{inflationfluctuation}):
\begin{equation}
\delta T_T=\delta T_A+\delta T_I
\end{equation}
and, therefore:
\begin{equation}
a_{\ell m}^T=a_{\ell m}^A+a_{\ell m}^I.
\end{equation}
Notice that, as discussed in \cite{inflationfluctuation}, 
there is the possibility that the 
inflation-produced contribution could be strongly biased or anti-biased
by the anisotropic background, mainly in the case in which
anisotropies grew as we go back in time. However as discussed 
in that reference, this is unlikely in general since it would require
a correlation between the quantum origin and subsequent classical evolution.
Moreover, in our case, the background evolution during inflation 
 is isotropic and we do not expect any interference effect. 

Following \cite{Campanelli}, we can easily generalize our results
to the case of an arbitrary orientation of our frame
in which the velocities lie
along the direction given by $(\hat\theta,\hat\phi)$. In that case, the
coefficients of the expansion are:
\begin{eqnarray}
&a_{20}^A&=\frac{\sqrt{\pi}}{6\sqrt{5}}[1+3\cos2\hat\theta]
\left|h_0-h_{dec}\right|,\nonumber \\
&a_{21}^A&=-(a_{2-1}^A)^*=-\sqrt{\frac{\pi}{30}}\,e^{-i\hat\phi}
\sin 2\hat\theta\left|h_0-h_{dec}\right|,\nonumber \\
&a_{22}^A&=(a_{2-2}^A)^*=\sqrt{\frac{\pi}{30}}\,e^{-2i\hat\phi}
\sin^2\hat\theta\left|h_0-h_{dec}\right|.
\end{eqnarray}
It is easy to show that the anisotropy quadrupole according to
(\ref{Qdefinition}) is still given by (\ref{qfe}) since $h$
is scalar under rotations. Now, assuming that the coefficients
$a_{2m}^I$ only differ one from each other in a phase factor we
can write:
\begin{eqnarray}
&a_{20}^I&=\sqrt{\frac{\pi}{3}}\,e^{i\alpha_1}Q_I,\nonumber\\
&a_{21}^I&=-(a_{2-1}^I)^*=\sqrt{\frac{\pi}{3}}\,e^{i\alpha_2}Q_I,\nonumber\\
&a_{22}^I&=(a_{2-2}^I)^*=\sqrt{\frac{\pi}{3}}\,e^{i\alpha_3}Q_I.
\end{eqnarray}
which is justified because the standard inflation fluctuations are
statistically isotropic. Then, the total quadrupole can be
expressed as:
\begin{equation}
Q^2_T=Q^2_A+Q_I^2-2fQ_AQ_I
\end{equation}
where $f$ is a function depending on the direction of the
velocities $(\hat\theta,\hat\phi)$ and the phase factors $\alpha_i$ of the
coefficients $a_{2m}^I$, and whose expression is:
\begin{eqnarray}
f=\frac{1}{4\sqrt{5}}\left[2\sqrt{6}\left[-\sin\hat\theta\cos(2\hat\phi
+\alpha_3)+2\cos\hat\theta\cos(\hat\phi+\alpha_2)\right]\sin\hat\theta
-(1+3\cos(2\hat\theta))\cos\alpha_1\right].
\end{eqnarray}
This function takes values such that:
\begin{equation}
|f|\leq \bar f=\frac{\sqrt{39+6\sqrt{6}}+\sqrt{6}-1}{4\sqrt{5}}.
\end{equation}
Since the values of the phases $\alpha_i$ are random, the total
quadrupole lies between $Q^2_+$ and $Q^2_-$, being:
\begin{equation}
Q^2_\pm=Q^2_A+Q_I^2\pm2\bar{f}Q_AQ_I
\label{ti}
\end{equation}
with $\bar{f}$ the maximum of $f$.

The observed quadrupole from WMAP \cite{WMAP3} is given by
$(\delta T)_{obs}^2 = 236^{+560}_{-137}\;\mu\K^2$ at the $68\%$ C.L.
or $(\delta T)_{obs}^2 = 236^{+3591}_{-182}\;\mu\K^2$ at the $95\%$ C.L. 
These results define the corresponding $68\%$ C.L or $95\%$ C.L. 
intervals for 
the measured temperature fluctuations that we denote: 
$((\delta T)^2_{min},(\delta T)^2_{max})$.
For the theoretical quadrupole temperature interval 
$((\delta T)^2_{-},(\delta T)^2_{+})$, obtained from (\ref{ti}),
to be compatible with observations,  we therefore require 
$(\delta T)^2_{max}\gsim(\delta T)^2_{-}$ and
$(\delta T)^2_{min}\lsim(\delta T)^2_{+}$. Using (\ref{ti})
these two conditions impose limits on $(\delta T_A)^2$ once the value of
$(\delta T)_I^2$ is fixed.

Let us first assume that 
inflation alone is able to account for the observed quadrupole, i.e., 
$(\delta T)_I^2\simeq 236\mu\K^2$,  then the first condition 
$(\delta T)^2_{max}\gsim(\delta T)^2_{-}$ is
automatically satisfied, because the minimum of $(\delta T)^2_+$, 
as a function of $\delta T_A$, is $(\delta T)_I^2$ which is
larger than $(\delta T)^2_{min}$. Therefore, we obtain bounds on
$\delta T_A$  just from the second condition above, which are given by:
\begin{eqnarray}
0\;\mu\K^2&\lsim& (\delta T_A)^2 \lsim 1861\;\mu\K^2 \;\;\ 68\% \;\mbox {C.L.}\nonumber \\
0 \;\mu\K^2&\lsim& (\delta T_A)^2 \lsim 5909\;\mu\K^2 \;\;\ 95\% \;\mbox {C.L.}
\label{excluded}
\end{eqnarray}

However, it is well-known that the predictions of standard inflation,  
calculated from an almost flat 
spectrum of density perturbations, is larger
than the central value of the measured quadrupole, in particular:
$(\delta T)_I^2\simeq 1252\mu\K^2$. In such a case the anisotropic 
contribution
could help reducing the value of the quadrupole for certain values of the 
phases and fluid velocities. Once again the first condition is
automatically satisfied,  and the second condition 
yields:
\begin{eqnarray}
54\;\mu\K^2\lsim(\delta T_A)^2\lsim
3857\;\mu\K^2 \;\;\ 68\% \;\mbox {C.L.}\nonumber \\
0\;\mu\K^2\lsim(\delta T_A)^2\lsim
9256\;\mu\K^2 \;\;\ 95\% \;\mbox {C.L.}
\label{constraint}
\end{eqnarray}
Notice that the $95\%$ confidence interval includes the 
standard prediction from inflation and for that reason the lower limit
vanishes in that case in (\ref{constraint}). According to these results, 
for certain orientations of the velocities and the
values of the phase factors, $Q_A$ could lower the value of 
quadrupole and make it compatible with the observed one 
even at the 1$\sigma$ level.
We will compare these limits with the predictions  from several
 models, but before that we need to extend our calculations
beyond the perturbative regime.

\section{Fast-moving fluids: exact equations}
In the previous Section we have studied the problem of obtaining
the quadrupole produced by the fact that dark energy does not share
a common rest frame with matter and radiation. To that end, we
have used cosmological perturbation theory to compute the
 metric perturbations by means of the
simple formula (\ref{hij}), valid  when the velocities are small.
Such formula could also be reasonably useful for high initial velocities
provided they drop in time and rapidly reach the perturbative regime.
If we look at Eq. (\ref{vevol}) we conclude that this
condition is satisfied if $w_\alpha<\frac{1}{3}$. Besides, if
$w_\alpha=\frac{1}{3}$, as in radiation case, the velocity is
constant so we just need to have a small initial velocity.
However, some models have been proposed
in which the total energy density of the universe
could contain in certain epochs a non-negligible
contribution of fluids with equation of state
such that $w_{\alpha}>\frac{1}{3}$. This is
for instance the case of stiff-fluid cosmologies or
some tracking dark energy models. In those cases  the velocities grow
in time, perturbation theory will eventually breaks down at some point
and  it becomes
 necessary to solve the exact problem.

In order to simplify the equations in this case, we shall
change the gauge used in the previous Section by one in which
$g_{00}=1$, keeping the condition $g_{0i}=0$. Moreover, we shall
also assume that the fluids are moving along the $z$-axis with no
rotation which means that axisymmetry holds. The most general
metric having that symmetry with this gauge choice can be written
as follows:
\begin{eqnarray}
ds^2=dt^2-a_{\perp}^{2}(dx^2+dy^2)-a_{\parallel}^2dz^2\label{am}
\end{eqnarray}
On the other hand, the
energy-momentum tensor for each fluid reads:
\begin{eqnarray}
T^{\,0}_{\alpha\,
0}&=&\gamma^2_\alpha\left(\rho_\alpha+p_\alpha\right)-p_\alpha,\nonumber\\
T^{\,i}_{\alpha\,
0}&=&\gamma_\alpha^2\left(\rho_\alpha+p_\alpha\right)v_{z\alpha}\delta^{iz},\nonumber\\
T^{\,0}_{\alpha\,
i}&=&-\gamma^2_\alpha\left(\rho_\alpha+p_\alpha\right)a_{\parallel}^2
v_{z\alpha}\delta_{i z},\nonumber\\
T^{\,i}_{\alpha\,
j}&=&-\gamma^2_\alpha\left(\rho_\alpha+p_\alpha\right)a_{\parallel}^2
v_{z\alpha}^2\delta^{i z}\delta_{j
z}-p_\alpha\delta^i_j.\label{aemt}
\end{eqnarray}
Note that the velocities appearing in these expressions are no
longer the same as those of the previous Section since, here, we
have defined them as derivatives with respect to the time $t$ not
with respect to $\eta$. However, it is easy to translate these
velocities into the others just by defining a mean scale factor
$a\equiv\sqrt[3]{a_{\perp}^2a_{\parallel}}$ because $dt\simeq a
d\eta$ and, therefore, $\frac{dx^i}{d\eta}\simeq
a\frac{dx^i}{dt}$, which is a good approximation in
the perturbative regime. Moreover, we have to notice that for an appropriate
definition of fluid velocity, we have to rescale
$V\equiv a_{\parallel}v$, so that $V^2\leq 1$.

Now, it is convenient to introduce the variables $\theta_\alpha$
defined by $\cosh \theta_\alpha=\gamma_\alpha$ so that the
equations adopt a simpler structure. Velocities are related to
$\theta_\alpha$ by means of
$\tanh\theta_\alpha=a_{\parallel}v_\alpha$. With these new
variables, the Einstein equations from (\ref{am}) and (\ref{aemt})
take the form:
\begin{eqnarray}
{ H}^2_\perp+2{H}_\perp{H}_\parallel &=&
8\pi G\sum_\alpha\left(\cosh^2\theta_\alpha
+w_\alpha\sinh^2\theta_\alpha\right)\rho_\alpha,\label{Hperp}\\
\dot{ H}_\perp
+\dot{ H}_\parallel+{ H}_\perp^2+{ H}_\parallel^2
+{ H}_\perp{ H}_\parallel &=&-8\pi G\sum_\alpha p_\alpha,\\
2\dot{ H}_\perp+3{
H}_\perp^2&=&-8\pi G\sum_\alpha\left(w_\alpha\cosh^2\theta_\alpha
+\sinh^2\theta_\alpha\right)\rho_\alpha,
\end{eqnarray}
where $\dot{}\equiv\frac{d}{dt}$ and ${
H}_\perp\equiv\dot{a}_\perp/a_\perp$, ${
H}_\parallel\equiv\dot{a}_\parallel/a_\parallel$ are the
transverse and longitudinal expansion rates respectively. These
equations reduce to the Friedmann ones when $a_\perp=a_\parallel$
and $v_{z\alpha}=0$. Again, as in the perturbative case, we need
some extra equations to close the problem which are those given by
the independent energy-momentum tensor conservation. These
equations can be written as follows:
\begin{eqnarray}
\dot{v}_{z\alpha}&=&-\frac{\left((w_\alpha-1)\cosh^2\theta_\alpha-1\right){
H}_\parallel+2w_\alpha{
H}_\perp}{(w_\alpha-1)\cosh^2\theta_\alpha-w_\alpha}\;v_{z\alpha},\label{eec}\\
\dot{\rho}_\alpha&=&\frac{(1+w_\alpha)\left({ H}_\parallel+2{
H}_\perp\cosh^2\theta_\alpha\right)}{(w_\alpha-1)\cosh^2\theta_\alpha-w_\alpha}
\;\rho_\alpha.
\end{eqnarray}
Besides, one can find the following equations for the evolution of
$\theta_\alpha$:
\begin{equation}
\dot{\theta}_\alpha=-\frac{(w_\alpha-1){ H}_\parallel+2w_\alpha{
H}_\perp}{(w_\alpha-1)\cosh^2\theta_\alpha-w_\alpha}\sinh\theta_\alpha\cosh\theta_\alpha.
\end{equation}
The spatial geodesic equations for the metric considered are:
\begin{eqnarray}
\frac{d^2x}{d\lambda^2}+2H_\perp \frac{dt}{d\lambda}\frac{dx}{d\lambda}=0,\nonumber\\
\frac{d^2y}{d\lambda^2}+2H_\perp\frac{dt}{d\lambda}\frac{dy}{d\lambda}=0,\nonumber\\
\frac{d^2z}{d\lambda^2}+2H_\parallel\frac{dt}{d\lambda}\frac{dz}{d\lambda}=0,
\end{eqnarray}
where $\lambda$ is an affine parameter. The first integral for
these equations is given by:
\begin{eqnarray}
\frac{d\vec{r}}{d\lambda}
=\left(\frac{n_x}{a_\perp^2},\frac{n_y}{a_\perp^2},\frac{n_z}{a_\parallel^2}\right)
\end{eqnarray}
being $\vec{r}=(x,y,z)$ and the integration constants
can be chosen for simplicity in such a way that $\vec n^2=1$.
 Moreover, from the condition of null
geodesic we get:
\begin{eqnarray}
\frac{dt}{d\lambda}=\sqrt{\frac{n_\bot^2}{a_\perp^2}+\frac{n_\|^2}{a_\parallel^2}}
\end{eqnarray}
with $n_\bot^2=n_x^2+n_y^2$ and $n_\|^2=n_z^2$. Then, for an
observer with velocity $u^\mu=\gamma(1,\vec v)$ the energy of the
photon is:
\begin{equation}
{\cal E}=\gamma
E\left[\sqrt{\frac{n_\bot^2}{a_\perp^2}+\frac{n_\|^2}{a_\parallel^2}}-\vec{n}\cdot\vec{v}\right]
\end{equation}
and the corresponding temperature fluctuation reads:
\begin{equation}
\frac{\delta
T}{T}=\frac{\gamma_0\,a_0}{\gamma_{dec}\,a_{dec}}\frac{\sqrt{\frac{n_\bot^2}{a_{\perp
0}^2}+\frac{n_\|^2}{a_{\parallel
0}^2}}-\vec{n}\cdot\vec{v}_0}{\sqrt{\frac{n_\bot^2}{a_{\perp
dec}^2}+\frac{n_\|^2}{a_{\parallel
dec}^2}}-\vec{n}\cdot\vec{v}_{dec}}-1
\end{equation}
where again the indices $0$ and $dec$ denote the present and decoupling
times respectively.

\section{Model examples}

\subsection{Constant equation of state}
The simplest dark energy model  we will consider is
that corresponding to a fluid with constant equation of state
$w_{DE}\simeq -1$. Note that in the case $w_{DE}=-1$, i.e. pure
cosmological constant, dark energy does not
contribute to the center of mass velocity (\ref{ccmc}) and, therefore, the center of mass
frame agrees with the radiation frame. This means that all the fluids
would share a common rest frame and no effects on the CMB would be
possible. When
$w_{DE}$ is close to $-1$, the velocity of dark energy
scales as $\sim a^{-4}$
and its energy density is nearly constant (see Fig. 1). Since dark
energy velocity decreases very fast, its contribution to the
quadrupole is very small. Besides, the velocity of radiation (and
therefore that of matter) is determined by the initial dark energy velocity
and the gauge condition (\ref{gaugecondition}) as:
\begin{equation}
\vec{v}_R=\frac{1+w_{DE}}{1+w_R}\frac{\Omega_{DE}}{\Omega_{R}}a_*^4
\vec{v}_{DE}^*.
\end{equation}
where $\vec v^*_{DE}$ is the initial dark energy velocity. Taking
$\Omega_{DE}=0.73$, $\Omega_{R}=8.18\times 10^{-5}$,
$w_{DE}=-0.97$ and $a_*\sim 10^{-6}$, we get
$\vec{v}_R\simeq2\times 10^{-22}\vec{v}^*_{DE}$. The value of
$a_*$ taken corresponds to a favorable case since lower values would
lead to much lower velocities of radiation and matter. Then, even
for initial velocities of dark energy close to 1, the velocities
of matter and radiation are extremely small, which means that the
three fluids are very nearly at rest in the cosmic center of mass
frame. That way, the quadrupole generated in this model is totally
negligible.

\begin{figure}[h]
{\epsfxsize=17 cm\epsfbox{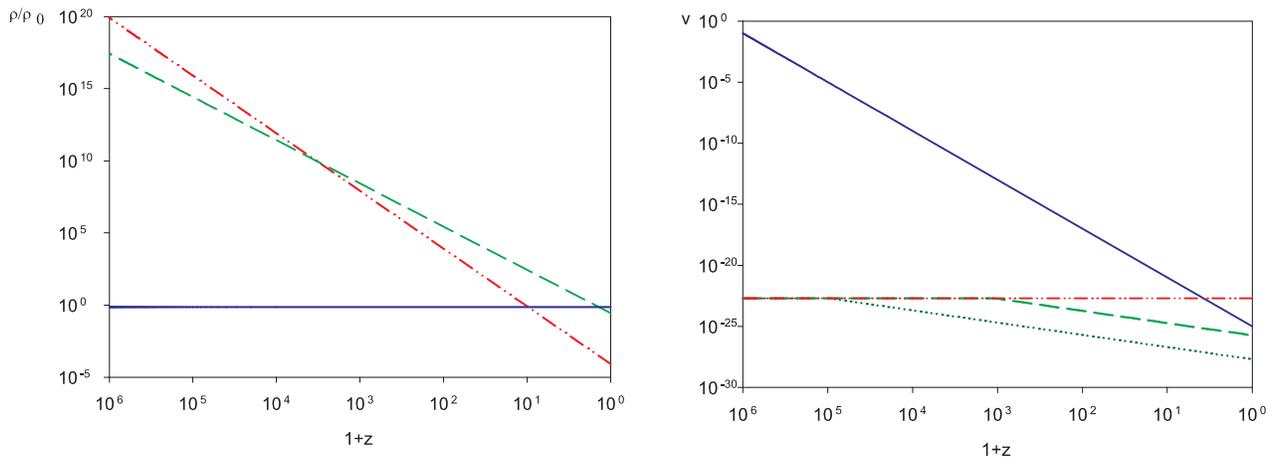}}
\caption{\small Evolution of
densities and velocities in a  model with constant equation of
state as described in the text.
Continuous line (blue) for dark energy, dashed-dotted (red)
for radiation, dotted (cyan) for dark matter and
dashed (green) for baryonic matter. On the left baryonic and dark matter are
added together and plotted in dashed (green). Notice that in this plot
dark matter is assumed to
decouple at $z\simeq 10^5$.}
\end{figure}


\subsection{Scaling models}
Scaling models \cite{scaling} are those with equation of state
such that dark energy mimics the dominant component of the
universe throughout most of the universe evolution. Thus, dark
energy evolves as radiation before matter-radiation equality and
as matter after that. However, in order to explain the accelerated
expansion of the universe, dark energy has to exit from that
regime and join into one with $w_{DE}<-1/3$ at some point. Then,
the evolution of the dark energy density is given by:
\begin{eqnarray}
\rho_{DE}=\left\{\begin{array}{c}
\rho_{DE\,0}\;a_{T}^{-3w_{DE}}a_{eq}a^{-4}
\;\;\;\;\;\;\;\;\;\;\;\;\;a<a_{eq}\\\\\\
\rho_{DE\,0}\;a_{T}^{-3w_{DE}}a^{-3}
\;\;\;\;\;\;\;\;\;a_{eq}<a<a_{T}\\\\\\
\rho_{DE\,0}\;a^{-3(w_{DE}+1)}
\;\;\;\;\;\;\;\;\;\;\;\;\;\;\;\;\;a>a_{T}\end{array} \right.
\end{eqnarray}
where as commented before, $a_{T}$ is the scale factor when
dark energy leaves
the scaling regime and $\rho_{DE\,0}$ is the present value of the
dark energy density.

In the evolution of dark energy velocity, we have to take into
account the momentum conservation equation given (to first order)
by (\ref{ecv1}). This equation implies that the dark
energy velocity must be discontinuous at the transition points since
the equation of state  jumps at those times whereas  the quantity
$a^4(1+w_{DE})\rho_{DE}\vec{v}_{DE}$ is constant, being
$\rho_{DE}$ continuous. With this in mind, we get the following
evolution for dark energy velocity:
\begin{eqnarray}
\vec{v}_{DE}=\left\{\begin{array}{c}
\vec{v}^*_{DE}
\;\;\;\;\;\;\;\;\;\;\;\;\;\;\;\;\;\;\;\;\;\;\;\;\;\;\;\;\;\;\;\;a<a_{eq}\\
\\\frac{4}{3}a_{eq}a^{-1}\vec{v}^*_{DE}\;\;\;\;\;\;\;\;\;\;\;
a_{eq}<a<a_{T}
\\\\
\frac{4a_{eq}a_{T}^{-3w_{DE}}}{3(1+w_{DE})}a^{3w_{DE}-1}\vec{v}^*_{DE}
\;\;\;\;\;\;a>a_{T}\end{array}
\right. .\label{vsm}
\end{eqnarray}

The discontinuities in the velocity arise because
we are considering  abrupt changes in the equation of state.
If these changes
were  smooth, the results would be  essentially unaffected
since the final values of the  velocities would remain those in
(\ref{vsm}). In Figure 2, we show the evolution of
the energy densities and velocities for a typical
scaling model.

We can see from the previous expression that, in the second
transition, the closer $w_{DE}$ is to $-1$, the more the velocity
grows after the transition. The case $w_{DE}=-1$ is not divergent
because, if that was the case, the conservation equation would
become trivial and the velocity evolution got from (\ref{vevol})
would not make sense anymore.
\begin{figure}[h]
{\epsfxsize=17 cm\epsfbox{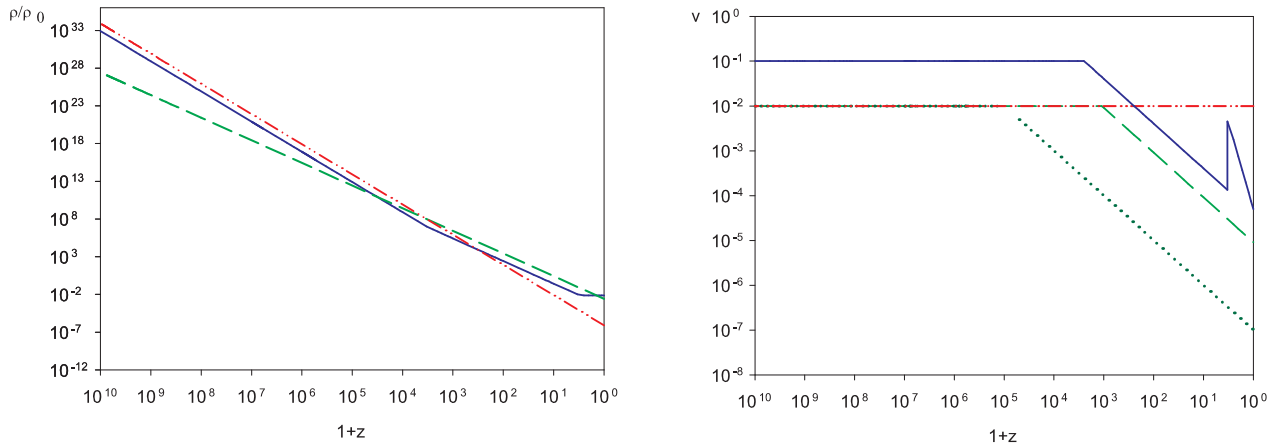}}
\caption{\small Densities and velocities evolution in a scaling model
with $v^*_{DE}=0.1$ and $\epsilon=0.1$. As in the previous figure,
the continuous line (blue) is for dark energy, dashed-dotted (red)
for radiation, dotted (cyan) for dark matter and
dashed (green) for baryonic matter. On the left baryonic and dark matter are
added together and plotted in dashed (green). Notice that in this plot
dark matter is also assumed to
decouple at $z\simeq 10^5$ and $a_*=10^{-10}$.}
\end{figure}

 In these scaling models, the first transition can be set at the
 matter-radiation equality and the second one must be
chosen such that we get the observed dark energy density  today.
Moreover, the initial velocity of radiation, and therefore that of
matter, is fixed by the initial velocity of dark energy via the
gauge condition (\ref{gaugecondition}).
Because matter is subdominant with respect to radiation before
equality,  the
matter contribution in (\ref{gaugecondition}) can be neglected and
we obtain that:
\begin{eqnarray}
\vec{v}_R^*=-\epsilon\vec{v}_{DE}^*\label{velrel}
\end{eqnarray}
where $\epsilon\equiv\rho_{DE}(a_*)/\rho_{R}(a_*)$ is the initial
dark energy density fraction (neglecting the matter contribution).
Notice that this fraction does not depend on $a_*$, because dark
energy scales as radiation in the radiation dominated era. Then,
we can obtain a relation between $a_{T}$ and $\epsilon$ just by
computing that quotient from the known expressions for the energy
densities evolutions of each fluid. When doing that it results:
\begin{equation}
a_{T}=\left[\frac{\Omega_{DE}
a_{eq}}{\Omega_R\epsilon}\right]^{\frac{1}{3w_{DE}}}.\label{aLM}
\end{equation}

Since we need $w_{DE}(z_T)<-1/3$ in order to have accelerated
expansion, we see from the previous formula that $a_{T}$ grows as
$\epsilon$ grows, more precisely if we take $w_{DE}(z<z_T)=
-0.97$, then $a_{T}\propto \epsilon^{0.34}$. Since primordial
nucleosynthesis imposes an upper limit on $\epsilon$, we can
establish also an upper limit on $a_{T}$ just by setting the
maximum value of $\epsilon$ on (\ref{aLM}). This maximum value is
$\epsilon_{max}\simeq 0.2$, (see for instance \cite{Kolb})
so we get the constraint $a_{T}\lsim
0.41$ for this kind of scaling models where we have taken
$\Omega_{DE}=0.73$, $\Omega_R=8.18\times10^{-5}$ and
$a_{eq}=\frac{1}{3300}$.
\begin{figure}[h]\begin{center}
{\epsfxsize=10.0 cm\epsfbox{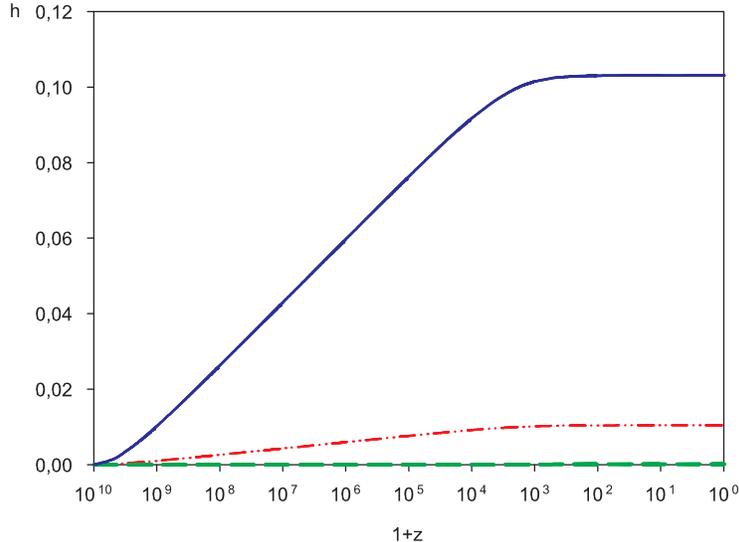}}
\caption{\small Evolution of metric perturbations due to each fluid in a
scaling model with $\epsilon=0.1$ and $v_*=0.1$. We can see that
the largest contribution comes from dark energy (continuous blue line)
and matter (dashed green) essentially does not contribute.
Radiation is shown with a dashed-dotted (red) line.
}\end{center}
\end{figure}

By inserting the corresponding values for the densities and
velocities of the four fluids, as well as the equation of state
considered into (\ref{halpha}), we can compute the quadrupole
produced by the relative motion of the fluids. For our
calculations we shall take $\Omega_B=0.046$, $\Omega_{DM}=0.23$,
$a_{dec}=\frac{1}{1100}$ and $w_{DE}(z<z_T)=-0.97$ and the values
given above. Fig. 3 shows the evolution of the contribution of
each fluid to the metric perturbation $h_\alpha$. We can see that
the typical behavior is a rapid growth during the radiation era to
reach finally a slightly growing regime in the matter era (notice that
the dependence on $a_*$ is only logarithmic). In
spite of the fact that the perturbation is $\Od(v_{DE}^{*\;2})$, 
the quadrupole is expected to be smaller because $h$ barely grows
in the epoch since decoupling to today and, as we mentioned in
Section 1, the quadrupole is essentially given by the growth of
the perturbation during that epoch.

The quadrupole produced by scaling models is fixed by two
parameters: the initial velocity $v_{DE}^*$  and
the initial energy ratio $\epsilon$ of dark energy. It is easy to
see from (\ref{halpha}) that $h_\alpha$ and, therefore the
quadrupole, is proportional to $v_{DE}^{*\;2}$. Obviously, this dependence
is valid just for small velocities since when we consider
velocities close to the speed of light we have to take into
account relativistic
effects. The dependence of the quadrupole on $\epsilon$ can be found to be
linear  for $\epsilon\lsim 0.07$ with a slope $0.44$ so we
can conclude that the quadrupole is very well approximated  by
the simple expression:
\begin{equation}
Q_A\simeq 0.44\,\epsilon\, v_{DE}^{*\;2}.
\label{QA}
\end{equation}
As commented before, this expression is valid only for small
velocities. According to the bounds on the quadrupole obtained in
(\ref{excluded}) and (\ref{constraint}), there are  allowed 
regions in the parameter space ($\epsilon, v_{DE}^*$), 
which from  (\ref{QA}) are limited by the curves
$\epsilon=k_\pm/v_*^2$ where the constants $k_\pm$ correspond
 to the
upper and lower limits on $Q_A$. In Fig. 4 we show these  
 regions
obtained numerically with the exact equations (see Section 4). As
we said above, the second order calculation is a good
approximation for velocities lower than $0.1$. However, when the
velocities are large (close to $1$) values of $\epsilon$
 $\lsim \Od(10^{-6})$ are necessary in order to explain the
observed quadrupole. Notice once again that these  regions have been
obtained in the case in which the measured quadrupole has two
contributions, one coming from inflation 
 and a second contribution coming from the fluids
motion.

\begin{figure}[h]
\begin{center}{\epsfxsize=10.0 cm\epsfbox{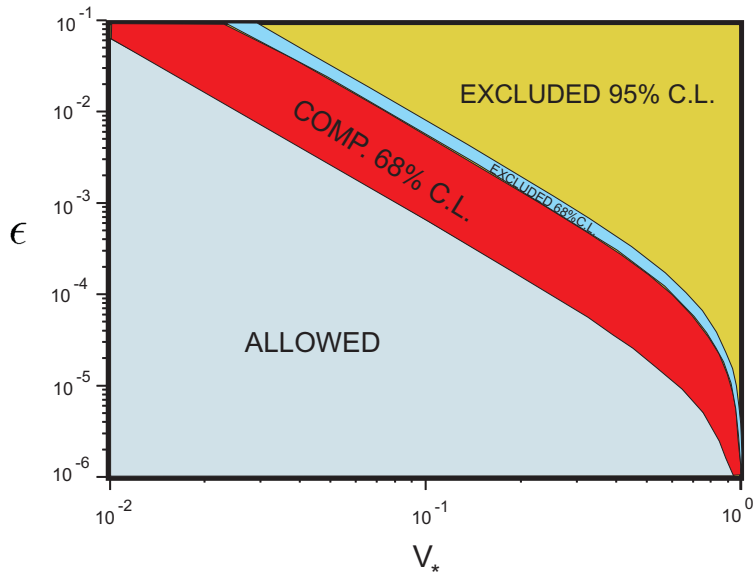}}
\caption{\small Exclusion plot in the parameter space
($\epsilon, v_{DE}^*$) for a  scaling dark energy model. The
allowed region corresponds to the limits given in (\ref{excluded}). 
The dark (red) strip corresponds to
the regions for which $Q_A$ could explain the observed quadrupole
at the 68$\%$ C.L. according to (\ref{constraint})}
\end{center}
\end{figure}

To end this section we shall show why we can neglect the third
term in (\ref{Tf}) with respect to that containing the metric
perturbation $h_{ij}$. Let us recall that term:
\begin{equation}
(\vec{v}\cdot\vec{n})|^0_{dec}(\vec{v}_{dec}\cdot\vec{n})\label{Qv}.
\end{equation}
The first factor in this expression is nothing but the dipole
which is $\sim 10^{-3}$. The second factor contains the
velocity of the observer at decoupling time which coincides with
matter velocity (and therefore with that of radiation) at that
moment. Then, if we recall the relation (\ref{velrel}) between
radiation and dark energy velocities, we find that this term is
$\sim 10^{-3}\epsilon\, v^*_{DE}$. On the other hand, the last term in
(\ref{Tf}) is $\sim \epsilon\, v_{DE}^{*\;2}$ as we have just seen above.
Hence, if we call $Q_v$ and $Q_h$ to the last two terms in
(\ref{Tf}) respectively we have that $Q_h\sim 10^3\,v^*_{DE}\,Q_v$ and we
see that $Q_v$ will be larger than $Q_h$ only for $v^*_{DE}< 10^{-3}$.
However, in
such a case the contribution to the quadrupole
is $\lsim 10^{-6}\epsilon$ which is  negligible.

\subsection{Tracking models}
In this Section we would like to comment on the difficulties which
can appear in certain dark energy models when we consider
perturbations in the fluids velocities. In general, any model with
a stiff stage in which its equation of state satisfies $w>\frac{1}{3}$, 
would be unstable with respect to velocity perturbations according
to (\ref{vevol}). This could be the case of certain tracking
models \cite{tracking}. These are models in which  the  energy
density of dark energy follows a common evolutionary track for a
wide range of initial conditions. This attractor behavior makes
this kind of models an interesting alternative to a cosmological
constant since they alleviate the so called coincidence problem.
Unlike scaling models, in this case dark energy does not
necessarily mimics the dominant component. In the model proposed
in \cite{tracking} the equation of state is initially  close to
1, then it
changes to $-1$ and, finally, it oscillates around $-0.2$.
\begin{figure}[h]
{\epsfxsize=17 cm\epsfbox{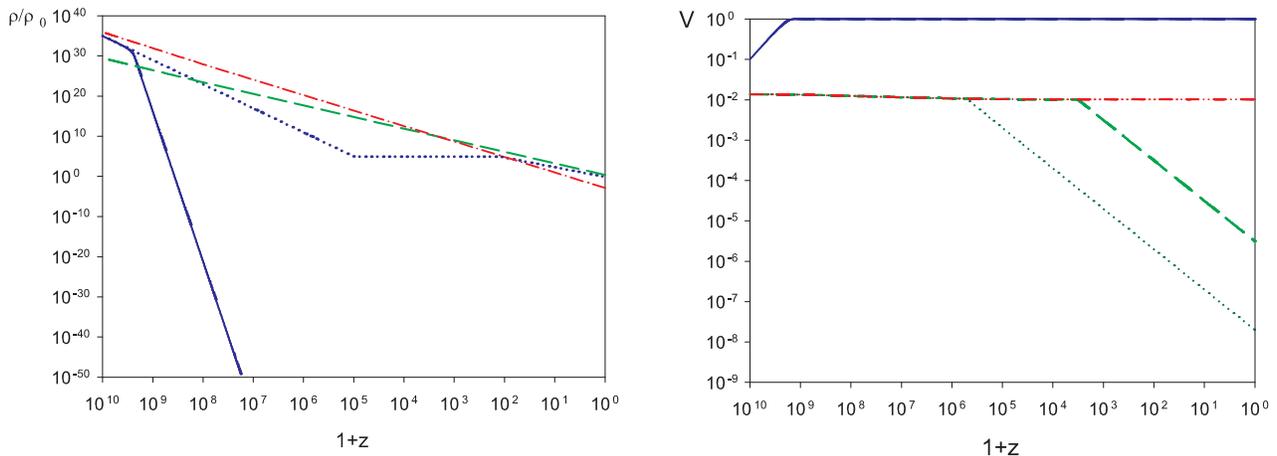}}
\caption{\small Densities and
velocities evolution in a typical tracking model with an initial 
equation of state $w_{DE}=0.9$, which changes to $w_{DE}=-1$ and then 
to $w_{DE}\sim -0.2$.
The continuous line (blue) is for dark energy in the model with
moving dark energy, whereas the blue dotted line is for
static dark energy, dashed-dotted (red)
for radiation, dashed (green) for baryonic and dark matter. 
 We
see that when $V_{DE}$ reaches $1$, the corresponding density
begins falling too fast to be able to recover the present value for
$\Omega_{DE}$.}
\end{figure}
Fig. 5 shows a typical behavior when $w>\frac{1}{3}$: first the
velocity perturbation (defined as $V\equiv a_\parallel v$) grows
according to (\ref{vevol}) and asymptotically  approaches  1. We
can understand this from the exact conservation equations
(\ref{eec}) by taking the ultrarelativistic limit $\theta\gg 1$.
This yields the solutions:
\begin{eqnarray}
v&=&v_0a_\parallel^{-1},\nonumber\\\rho&=&\rho_0a_\perp^{-2\frac{1+w}{1-w}}.
\end{eqnarray}
This means that $V\simeq 1$  is a solution of the equations. In addition, 
the energy density for $w> 1/3$ falls very fast with the expansion
when compared with the usual behaviour $\rho=\rho_0 a^{-3(1+w)}$.
In addition,  $\gamma^2\rho$, which is the quantity 
that contributes
to the Hubble rate in (\ref{Hperp}), decays as $(a_{\parallel}
a_{\perp})^{-2}$, once the fluid reaches the ultra-relativistic regime, 
regardless
the value of $w$.

In the limiting case of stiff fluids with
$w=1$, it is possible to obtain exact solutions. Thus
 the velocity perturbation and the energy density  are:
\begin{eqnarray}
V&=&V_0a_\perp^2,\nonumber\\\rho&=&\rho_0\frac{e^{-4\int{H_\perp\cosh^2\theta
dt}}}{a_\parallel^2}.
\end{eqnarray}

Thus, the velocity of the fluid grows as $a_\perp^2$ until it reaches
the speed of light in a finite
time and the density falls to zero at the same time because
$\theta$ becomes infinity at that moment. From that time on,
the fluid will keep moving at the speed of light with
vanishing energy density. Notice that $\gamma$ becomes
infinity in such a way that the momentum is conserved i.e. the
mathcing between the two regimes must be taken
so that $\gamma^2\rho$ is finite and
continuous. In any case, we see that when the velocity is high
enough the density
falls to zero and, the closer is $w$ to one, the faster is the
fall of the density, so that we cannot recover the present value for
the dark energy density. Notice that this general behaviour is
independent of the value of the initial velocity and, accordingly,
even in models in which all the fluids are initially at rest, a small
perturbation in the velocity could change dramatically the final values of
the densities, unless fine tunings of the transition redshifts
are introduced.

\subsection{Null dark energy}
In this Section we shall study the case in which dark energy
behaves as a null fluid, whose energy-momentum tensor reads:
\begin{equation}
T^{\mu\nu}_N=(\rho_{N}+p_{N})l^\mu l^\nu-p_Ng^{\mu\nu}\label{NF}
\end{equation}
with $l^\mu$ a null vector given, in the Bianchi type I metric
(\ref{am}), by $l^\mu=(1,0,0,a_\parallel^{-1})$. For this kind of
fluid, the conservation of energy and momentum can be expressed as
follows:
\begin{eqnarray}
0&=&\dot{p}_N,\\
0&=&(\dot{\rho}_N+\dot{p}_N)+2(H_\parallel+H_\perp)(\rho_N+p_N).
\end{eqnarray}
These equations imply that the pressure is constant and that
$(\rho_N+p_N)$ scales as $(a_\parallel a_\perp)^{-2}$, so that the
energy density is given by $\rho_N=\rho_{N0}(a_\parallel
a_\perp)^{-2}-p_{N0}$ where $p_{N0}$ and $\rho_{N0}$ are
constants of integration. Since the anisotropy is expected to be
small, the energy density of this fluid behaves as radiation
during the early epoch and as a cosmological constant with energy
density
$-p_{N0}$ at late times. Now, if we require $\rho_N$ to be
positive at all times, we conclude that the pressure must be negative, as
corresponds to a cosmological constant.
Notice that this is a general result for any null fluid whose
energy-momentum
tensor is
given by (\ref{NF}). The transition between
both regimes can be easily calculated and it is given by
$a_T\simeq (-\frac{\rho_{N0}}{p_{N0}})^{1/4}$. Since
$p_{N0}=-0.73\rho_{c0}$, where
$\rho_{c0}$ is the critical density today,  we
have that $a_T=0.1\epsilon^{1/4}$ where
$\epsilon\equiv\frac{\rho_{N0}}{\rho_{R0}}$ is the ratio of dark
energy density with respect to radiation  which  is
almost constant. Besides, this ratio is also the
initial contribution of dark energy to the total energy density
which has an upper limit imposed by primordial nucleosynthesis.
Taking once again $\epsilon_{max}\lsim 0.2$,
we have an upper limit on the
transition  given by: $a_T\lsim 2\times 10^{-2}$.

The exact Einstein equations in this case are:
\begin{eqnarray}
{H}^2_\perp+2{H}_\perp{H}_\parallel &=&
8\pi G\sum_\alpha\left(\cosh^2\theta_\alpha
+w_\alpha\sinh^2\theta_\alpha\right)\rho_\alpha+8\pi G\rho_N,\\
\dot{ H}_\perp+\dot{ H}_\parallel+{ H}_\perp^2+{ H}_\parallel^2
+{ H}_\perp{ H}_\parallel &=&-8\pi G\sum_\alpha p_\alpha-8\pi G p_N,\\
2\dot{ H}_\perp+3{
H}_\perp^2&=&-8\pi G\sum_\alpha\left(w_\alpha\cosh^2\theta_\alpha
+\sinh^2\theta_\alpha\right)\rho_\alpha-8\pi G(\rho_N+2p_N),\label{EENF}
\nonumber\\
\end{eqnarray}
where now $\alpha=B,DM,R$.  Moreover, we still have the gauge
condition $\vec{S}=0$ which yields the following constraint:
\begin{equation}
\sum_\alpha\gamma^2_\alpha(\rho_\alpha+p_\alpha)v_\alpha+(\rho_N+p_N)=0.
\end{equation}
In the radiation-dominated era we can neglect the contribution of
matter (dark matter and baryons) to the latter sum, so we get:
\begin{equation}
\gamma_R^2v_R=-\frac{\rho_N+p_N}{\rho_R+p_R}\label{GCNF}
\end{equation}
Since, in that epoch, dark energy must be subdominant with respect
to radiation, the quotient on the RHS is small and, therefore, the
velocity of radiation is also small. This allows us to consider
the perturbative regimen in the velocities (except, obviously for the
null fluid).

Therefore, if we assume that the anisotropy generated is small we can
set the following form for $a_\parallel$ and $a_\perp$:
\begin{eqnarray}
a_\perp&=&a(1+\delta_\perp),\nonumber\\
a_\parallel&=&a(1+\delta_\parallel).\label{aexp}
\end{eqnarray}
With this ansatz it is easy to see that
$h=2(\delta_\parallel-\delta_\perp)$. Then, inserting (\ref{aexp})
in (\ref{EENF}) and expanding up to first order in $\delta$'s and
$v_\alpha$ we can get the following equation for $h$:
\begin{equation}
\frac{d}{dt}\left(a^3\frac{dh}{dt}\right)=2a^3(\rho_N+p_N).
\end{equation}
This equation can be easily solved by means of two direct
integrations and its solution can be expressed as follows:
\begin{equation}
h=6\int_{a_*}^a\frac{1}{\tilde{a}^4}\left[\int_{a_*}^{\hat{a}}\hat{a}^2(\rho_N+p_N)\frac{d\hat{a}}{\sqrt{\sum_\alpha\rho_\alpha}}\right]\frac{d\tilde{a}}{\sqrt{\sum_\alpha\rho_\alpha}}.\label{nullh}
\end{equation}
In principle, the problem is not solved yet since $\rho_N+p_N$
depends on $\delta_\parallel$ and $\delta_\perp$. However, we can
consider the lowest order in this quantity, i.e.,
$\rho_N+p_N=\rho_{N0}a^{-4}$ to obtain the dominant contribution
to the quadrupole. This is justified because
$\frac{\rho_{N0}}{\rho_{R0}}$ is of the same order as $v_R$ as we
can see from (\ref{GCNF}). We have to note that, to this order,
the quadrupole depends just on the null fluid because the first
contribution to the anisotropy due to the rest of  fluids is of
second order in the velocities, whereas the null fluid contributes
to first order. In Fig. 6 we plot the evolution of the fluids densities
and $h$ function for a null fluid with $\epsilon=5\times 10^{-6}$.
\begin{figure}[h]
{\epsfxsize=17.0 cm\epsfbox{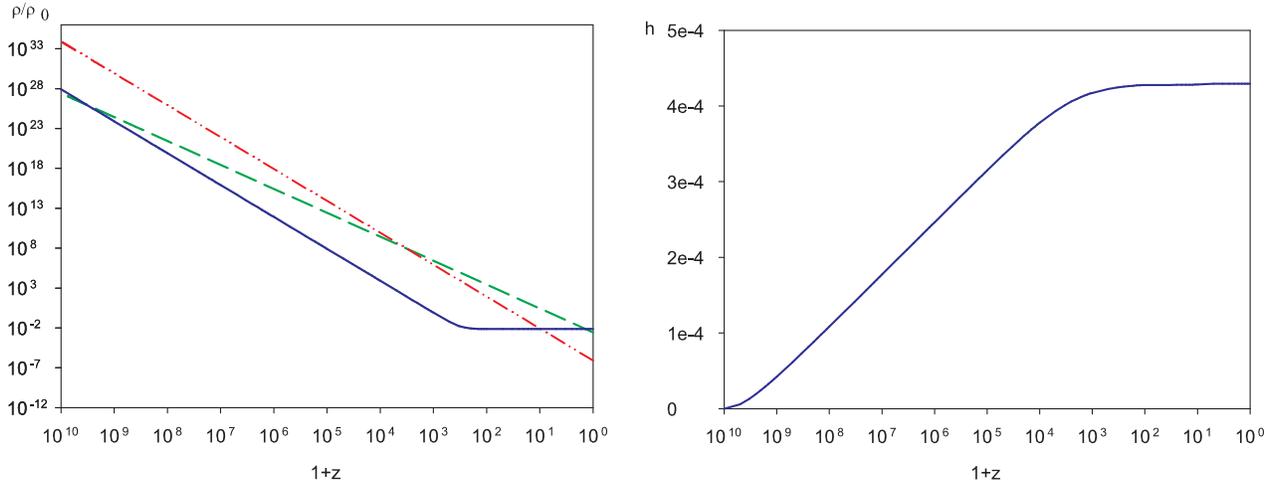}}
\caption{\small Left:
densities evolution for a null fluid with $\epsilon=5\times 10^{-6}$.
 Matter is plotted with dashed (green) line, dotted-dashed (red) for
radiation and continuous (blue) line for dark energy.
We see that the null fluid behaves as in a scaling model
except during the matter dominated epoch. Right:
evolution of $h$ showing that the anisotropy grows up to a maximum
value where it keeps almost constant.}
\end{figure}

In this model we only have one free parameter: $\rho_{N0}$ or,
equivalently, $\epsilon\equiv\frac{\rho_{N0}}{\rho_{R0}}$, so we
can get bounds on $\epsilon$ just from (\ref{excluded}) and (\ref{constraint}).
Besides, the quadrupole is linear in $\epsilon$, as we see looking at
(\ref{nullh}), more precisely we have that the quadrupole is given
by: $Q_A\simeq 2.58\,\epsilon$. Now again the contribution  from the
term  $Q_v$  given by (\ref{Qv}) is negligible compared with $Q_A$ since
$Q_v\simeq 10^{-3}\epsilon$.

This expression is nearly independent
of $a_*$ because the quadrupole depends on the difference
$h_0-h_{dec}$ which is not very sensitive to the time at which we
set the initial conditions. Comparing the expression obtained
for the quadrupole with the previous bounds, we
get that the allowed region in (\ref{excluded}) corresponds to:
\begin{eqnarray}
\epsilon&\lsim& 6.1\times 10^{-6} \;\;\; 68\%\;\mbox{C.L.}\nonumber \\
\epsilon&\lsim& 1.1\times 10^{-5} \;\;\; 95\%\;\mbox{C.L.}
\end{eqnarray}
whereas for:
\begin{eqnarray}
1\times10^{-6}\lsim\epsilon\lsim 8.8\times 10^{-6}\;\;\;
68\%\;\mbox{C.L.}\nonumber \\
0\lsim\epsilon\lsim 1.4\times 10^{-5}\;\;\;
95\%\;\mbox{C.L.}
\end{eqnarray}
the null fluid could make the predicted quadrupole to agree
with observations, as shown in (\ref{constraint}).

\section{Conclusions and discussion}

In this paper we have studied homogeneous models of dark energy
in which the rest
frame of the different fluids can differ from each other. We have
considered the evolution of slow-moving and fast-moving fluids and
shown that, starting from an initially isotropic universe, the fluids
motions can generate an anisotropic expansion in which the anisotropy
degree typically grows in time.
Such anisotropies are shown to contribute to the
CMB dipole and quadrupole only, but not to higher multipoles.
We apply those results to some dark energy models and find that in
models with constant equation of state, even for initial
velocities of dark energy close to the speed of light, throughout the
matter era
all the fluids would practically  share a common rest frame and therefore no
effect on the quadrupole is expected. However, in the case of scaling
models it is shown that the anisotropy grows during the radiation era
and reaches an almost constant value during matter domination.
The effect on the CMB quadrupole can be relevant and bounds on the velocity
and  initial fraction of dark energy can be found. We also find that
for  models with an initial stage in which the equation
of state is stiffer than radiation, as for instance in some tracking models,
the velocity approaches the speed
of light whereas the energy density decays faster than in the case in which
dark energy is at rest with respect to matter and radiation. 
This fact spoils the predictions of those models for the density
parameters at late times. Finally we have considered also fluids moving at
the speed of light and found that generically they behave as a cosmological
constant at late time, provided their energy density is positive at all times,
 whereas they act as radiation at early times. The contribution to the quadrupole is
also used to set limits on the relative contribution of dark energy
in the radiation dominated era.

The models presented in this paper with moving fluids lead to
Bianchi I anisotropic metrics of the type recently studied
in \cite{Campanelli}. However unlike that model
with decaying anisotropies generated by the presence of magnetic fields
in the early universe, in our work the motion of dark energy
supports the anisotropies which could have a non-negligible value today.

Finally concerning the potential effects on the CMB polarization, 
it is well known  \cite{Rees}
that Bianchi models give rise to polarized radiation
through Thomson scattering at decoupling time. This is also
the case during the reionization period and therefore a
potential contribution from dark energy motion is expected also in the
polarization signal at large scales. Work is in progress in this direction.

{\em Acknowledgments:} We would like to thank A. de la Cruz Dombriz for
usufel comments. This work has been partially supported by
DGICYT (Spain) under project numbers FPA 2004-02602 and FPA
2005-02327. The latter project has supported J.B. under the grant
BES-2006-12059 by Ministerio de Educaci\'on y Ciencia.

\end{document}